\documentclass[aps,prl
,twocolumn%
,superscriptaddress%
,showpacs]{revtex4-1}
\usepackage{graphicx}%
\usepackage{dcolumn}
\usepackage{amsmath}

\makeatletter
\def\btt#1{\texttt{\@backslashchar#1}}%
\DeclareRobustCommand\bblash{\btt{\@backslashchar}}%
\makeatother

\begin{document}
\bibliographystyle{apsrev}
\title{Unusual angular dependence of tunneling magneto-Seebeck effect} 
\author{Christian Heiliger}%
 \email{christian.heiliger@physik.uni-giessen.de}
\affiliation{%
I. Physikalisches Institut, Justus Liebig University, Giessen, Germany
}
\author{Michael Czerner}%
\affiliation{%
I. Physikalisches Institut, Justus Liebig University, Giessen, Germany
}
%
\author{Niklas Liebing}%
\affiliation{%
Physikalisch-Technische Bundesanstalt, Braunschweig, Germany
}

\author{Santiago Serrano-Guisan}
\affiliation{%
International Iberian Nanotechnology Laboratory, Av. Mestre Jos\`e Veiga, 4715-330 Braga, Portugal
}

\author{Karsten Rott}
\affiliation{%
Department of Physics, Universit\"at Bielefeld, 33501 Bielefeld, Germany
}

\author{G\"unter Reiss}
\affiliation{%
Department of Physics, Universit\"at Bielefeld, 33501 Bielefeld, Germany
}

\author{Hans W. Schumacher}%
\affiliation{%
Physikalisch-Technische Bundesanstalt, Braunschweig, Germany
}
\date{\today}
\begin{abstract}
We find an unusual angular dependence of the tunneling magneto-Seebeck effect (TMS). The conductance shows normally a cosine-dependence with the angle between the magnetizations of the two ferromagnetic leads. In contrast, the angular dependence of the TMS depends strongly on the tunneling magneto resistance (TMR) ratio. For small TMR ratios we obtain also a cosine-dependence whereas for very large TMR ratios the angular dependence approaches a step-like function.

\end{abstract}
%
\maketitle
%
%
%
Spin caloritronics \cite{bauer10,bauer12} combines the spin-dependent charge transport with energy or heat transport. Therefore, the spin degree of freedom is used in addition to thermoelectrics.
The basic physics was already pointed out by Johnson and Silsbee \cite{silsbee87}.
Recently a number of new effects were discovered and discussed like
spin-Seebeck effect \cite{uchida08,xiao10}, 
magneto-Seebeck effect in metallic multilayers \cite{gravier06},
tunneling magneto-Seebeck effect (TMS) \cite{czerner11,walter11,liebing11},
thermal spin-transfer torque \cite{jia11},
Seebeck spin tunneling \cite{lebreton11},
thermally excited spin-currents \cite{tsyplyatyev06},
magneto-Peltier cooling \cite{hatami09}. 

In this letter we focus on the recently experimentally observed tunneling magneto-Seebeck effect (TMS) \cite{walter11,liebing11} in MgO based tunnel junctions, which we predicted in our previous paper \cite{czerner11}. The tunnel junction we consider consists of two ferromagnetic leads that are separated by a MgO barrier. Thereby, $\theta$ is the angle between the magnetizations of the two ferromagnetic leads.
In our previous studies we found for the angular dependence of the Seebeck coefficient 
an almost step-like function based on \textit{ab initio} theory \cite{czerner11} whereas our previous experiments show a cosine-dependence \cite{liebing11}. Other transport quantities like conductance or spin-transfer torque show in general a cosine or sine dependence \cite{jaffres01,slonczewski89,heiliger08}. Therefore, the aim of this letter is to investigate the angular dependence of the TMS on general arguments and compare these findings to \textit{ab initio} calculations as well as experiments.

Starting point is the energy dependent transmission probability $T(E)$ from which the conductance $G$ and Seebeck coefficient $S$ 
\begin{equation}
G=e^2 L_0 \ \ \ \ \ \ S=-\frac{1}{e \Theta} \frac{L_1}{L_0}
\label{eq:G_S}
\end{equation}
can be calculated by the moments \cite{ouyang09}
\begin{equation}
L_n=\frac{2}{h} \int T(E) (E-\mu)^n (-\partial_E f(E,\mu,\Theta)) dE \ ,
\label{eq:L_n}
\end{equation}
where $f(E,\mu,\Theta)$ is the Fermi occupation function at a given energy $E$, electrochemical potential $\mu$, and temperature $\Theta$.
Usually, the transmission $T(E,\theta)$ at a given energy and a relative magnetic orientation between the two ferromagnetic layers shows a cosine dependence going from parallel (P) $T^P(E)=T(E,0^{\circ})$ to anti-parallel (AP) $T^{AP}(E)=T(E,180^{\circ})$ alignment of the ferromagnetic layers \cite{slonczewski89}
\begin{equation}
T(E,\theta) = \frac{T^P(E)+T^{AP}(E)}{2} + \frac{T^P(E)-T^{AP}(E)}{2} \cos (\theta) \ .
\label{eq:trans}
\end{equation}
Plugging this into Eqs. (\ref{eq:G_S}) and (\ref{eq:L_n}) leads to the cosine-like angular dependence of the conductance
\begin{equation}
G(\theta) = \frac{G^P+G^{AP}}{2} + \frac{G^P-G^{AP}}{2} \cos (\theta) \ \ .
\label{eq:cond}
\end{equation}
Consequently, the resistance $R(\theta)=1/G(\theta)$ is not cosine-like but actually depends on the size of the TMR ratio \cite{jaffres01}. Using the optimistic definition of the TMR ratio
\begin{equation}
\text{TMR} = \frac{G^P-G^{AP}}{G^{AP}}= \frac{R^{AP}-R^{P}}{R^{P}} 
\label{eq:TMR}
\end{equation}
one gets for the angular dependence of the resistance
\begin{equation}
R(\theta) = \frac{2 R^{AP}}{\text{TMR}+2} \ \ \frac{1}{1+\frac{\text{TMR}}{\text{TMR}+2} \cos (\theta)} \ \ .
\label{eq:R_theta}
\end{equation}
This means that a cosine-like dependence for the resistance is only approximately valid if $\frac{\text{TMR}}{\text{TMR}+2}$ is small, which means the TMR ratio is small.

Let us now turn to the TMS effect. Plugging Eq. (\ref{eq:trans}) into Eqs. (\ref{eq:G_S}) and (\ref{eq:L_n}) leads to the angular dependence of the Seebeck coefficient
\begin{equation}
S(\theta) = \frac{S^P G^P + S^{AP} G^{AP} + (S^P G^P - S^{AP} G^{AP}) \cos (\theta)}{G^P + G^{AP} + (G^P - G^{AP}) \cos (\theta)} \ \ .
\label{eq:S_theta}
\end{equation}
Using the definition of the TMR ratio in Eq. (\ref{eq:TMR}) we get
\begin{equation}
S(\theta) = \frac{S^P \text{TMR} + S^P + S^{AP} + (S^P \text{TMR} + S^P - S^{AP} ) \cos (\theta)}{\text{TMR} + 2 +\text{TMR} \cos (\theta)} \ \ .
\label{eq:S_theta_TMR}
\end{equation}

Equation (\ref{eq:S_theta_TMR}) is the main result of our paper. It shows that the angular dependence of the Seebeck coefficient is rather complex and depends on the TMR ratio. Therefore, it is worth to consider two special cases:

First, we want to consider a vanishing TMR ratio $\text{TMR}=0$. Then Eq. (\ref{eq:S_theta_TMR}) simplifies to
\begin{equation}
S(\theta) = \frac{S^P + S^{AP}}{2} + \frac{(S^P - S^{AP} ) \cos (\theta)}{2} \ \ ,
\label{eq:S_theta_TMR_0}
\end{equation}
which is the cosine-like angular dependence that we already obtained for the conductance in Eq. (\ref{eq:cond}). Another important result is evident: although the TMR vanishes a TMS effect can exist.

Second, we want to consider very large TMR ratios. For this purpose we rewrite Eq. (\ref{eq:S_theta_TMR}) to
\begin{equation}
S(\theta) = \frac{S^P + \frac{S^P + S^{AP}}{\text{TMR}} + (S^P + \frac{S^P - S^{AP}}{\text{TMR}} ) \cos (\theta)}{1+ \frac{2}{\text{TMR}} + \cos (\theta)} \ \ .
\label{eq:S_theta_TMR_inf}
\end{equation}
In the limit of an infinite TMR ratio $S(\theta<180^\circ)=S^P$. On the other hand we know that $S(\theta=180^\circ)=S^{AP}$. Consequently, $S(\theta)$ becomes a step-like function in the limit of infinity TMR ratio
\begin{equation}
\lim\limits_{\text{TMR} \rightarrow \infty} S(\theta)  = \begin{cases}
							S^P & \text{if } \ \theta < 180^\circ  \\
							S^{AP} & \text{if } \ \theta = 180^\circ 
            \end{cases} \ \ .
\label{eq:S_theta_TMR_step}
\end{equation}

\begin{figure}
\includegraphics[width=0.99 \linewidth]{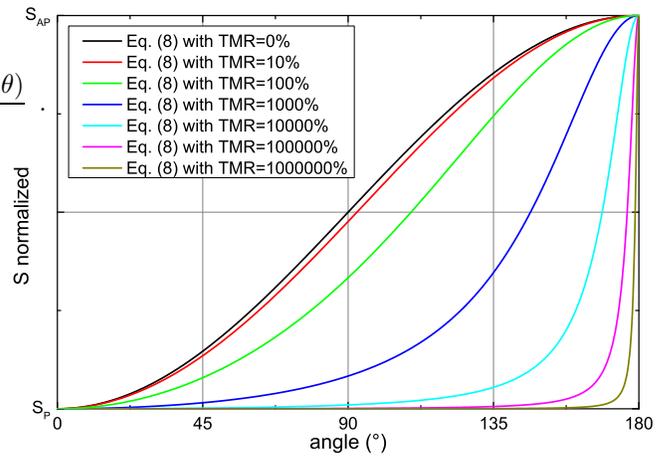}
\caption{
(Color online) Seebeck coefficient given by Eq. (\ref{eq:S_theta_TMR}) as a function of the angle $\theta$ between the magnetic orientations of the ferromagnetic leads for different TMR ratios.
}
\label{fig:model}
\end{figure}

Figure \ref{fig:model} shows the angular dependence of $S$ for TMR ratios between these both limiting cases. This viewgraph shows that the angular dependence of $S$ is strongly affected by the TMR ratios. Consequently, this variation with the TMR ratio can explain the different observations in the experiment and theory for MgO based tunnel junctions because in the experiment typically TMR ratios are about $100\%$ whereas in theory typically TMR ratios are several $1000\%$.

\begin{figure}
\includegraphics[width=0.99 \linewidth]{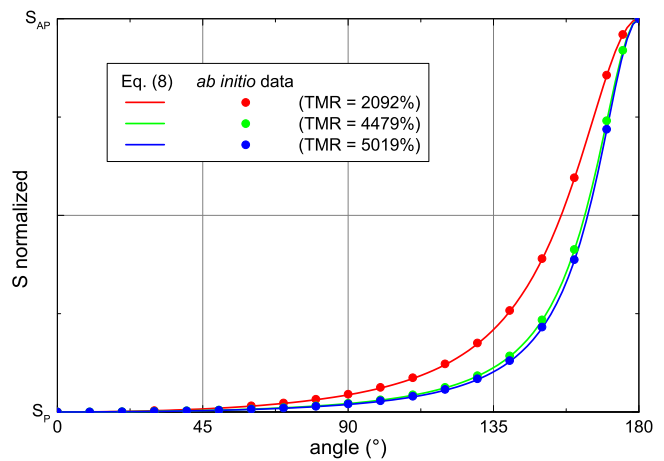}
\caption{
(Color online) Comparison of the angular dependence of the Seebeck coefficient between our previous \textit{ab initio} calculations \cite{czerner11,walter11,heiliger13} and Eq. (\ref{eq:S_theta_TMR}). The values are at 300K with a temperature gradient of 10K.
}
\label{fig:comp_theo}
\end{figure}

To validate Eq. (\ref{eq:S_theta_TMR}) we compare it in Fig. \ref{fig:comp_theo} to our previous \textit{ab initio} calculations \cite{czerner11,walter11,heiliger13}. The viewgraph clearly shows that Eq. (\ref{eq:S_theta_TMR}) is indeed valid and that we have a perfect match of the \textit{ab initio} results with the qualitative model given in Eq.~(\ref{eq:S_theta_TMR}). Consequently, our findings can explain the different observed angular dependence in theory and experiment due to different TMR ratios.

\begin{figure}
\includegraphics[width=0.99 \linewidth]{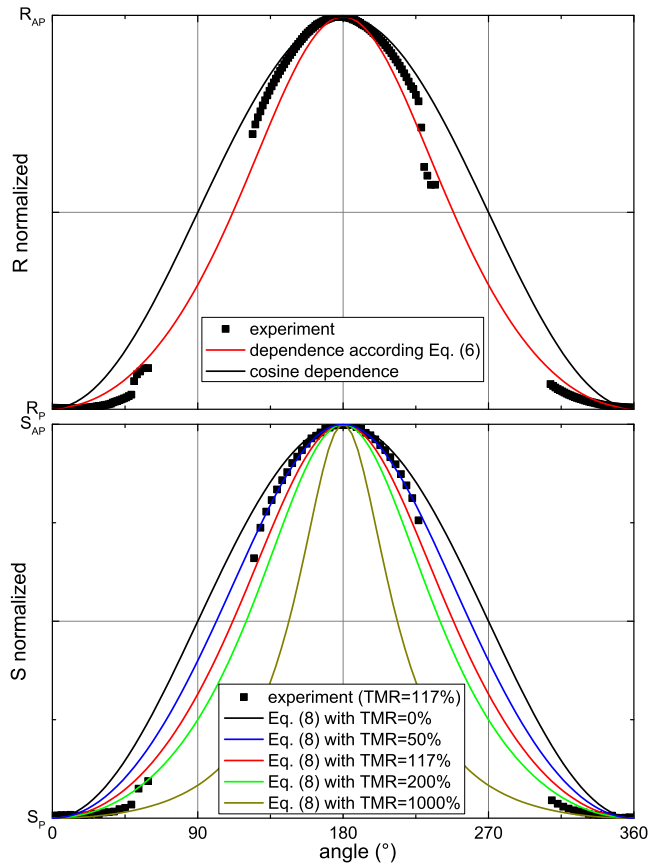}
\caption{
(Color online) Top: Comparison of the angular dependence of the resistance between our experimental data and Eq. (\ref{eq:R_theta}). In addition, we plot a fictitious cosine dependence. Bottom: Comparison of the angular dependence of the Seebeck coefficient between our experimental data and Eq. (\ref{eq:S_theta_TMR}) for different TMR ratios. The TMR ratio in the experiment is $117\%$. 
}
\label{fig:comp_exp}
\end{figure}

To experimentally test the predictions of Eq.~(\ref{eq:S_theta_TMR}) we compare in Fig.~\ref{fig:comp_exp} our findings to our experimental results obtained on a nanopillar fabricated from a CoFeB/MgO/CoFeB based magnetic tunnel junction. The sample structure and setup for the electrical measurements of the TMR and thermopower have been described in References \cite{serrano08} and \cite{liebing11}, respectively. For the angular dependent measurements the temperature gradient was created by direct current heating with heating power up to 110 mW to generate temperature gradients of a few tens of mK across the MgO barrier. The Oersted field generated by the heater line is carefully compensated by an external field onto which the rotational field is superimposed. A 30 mT vector field is rotated in steps of $4^\circ$ and the TMR and magneto thermopower are measured. Note that for the given vector field the free layer magnetization is not fully aligned to the field vector due to the influence of 
the effective anisotropy. The anisotropy function is determined by means of switching field angular measurements of the MTJ. The obtained critical curve of the free layer is used to calculate the magnetization direction of the free layer for each applied field vector \cite{thiaville98}.
In Fig.~\ref{fig:comp_exp} the experimental data of TMR (top) and the Seebeck coefficient (bottom) are plotted as function of the free layer angle for comparison to theory. 

In particular, we compare our experimental TMR and TMS data to Eq. (\ref{eq:R_theta}) and to Eq. (\ref{eq:S_theta_TMR}) respectively for different TMR ratios. The experimental TMR data shows mainly a single domain behavior and thus a smooth angular dependence with a TMR ratio of about 117 \% (see Fig.~\ref{fig:comp_exp} top). However, near $60^\circ$ and $240^\circ$ of the free layer magnetization steps of the TMR curves indicate the influence of small non-uniform domains near the switching across the hard axis. As a consequence of these steps a variety of theoretical curves with TMR ratios around 100 \% would fit the experimental angular dependence of TMR and TMS. Thus, fitting the experimental TMS data to Eq. (\ref{eq:S_theta_TMR}) a TMR ratio of 78 \% is derived, which is close to the measured TMR ratio in Fig.~\ref{fig:comp_exp} top keeping in mind the scattering of data points. Also, a deviation of the determined anisotropy function by the astroid method \cite{thiaville98} from the real, rather complex, effective anisotropy causes an error in the calculated free layer magnetization orientation and thus in the dependence of TMR and TMS curves. 
Nevertheless, there is a significant deviation from the cosine-like dependence in both TMR and TMS curves, as a consequence of the large measured TMR ratio, as pointed out previously in the text.

This is also an example how the size of the TMR ratio can be extracted from the qualitative angular dependence of the TMS. On a first view this is somehow surprising, because due to the definition given in Eq.~(\ref{eq:G_S}) there is no general relationship between the conductance and the Seebeck coefficient, unless for very low temperatures where a Sommerfeld expansion of the occupation function is possible.

In summary, we derived an expression (\ref{eq:S_theta_TMR}) for the angular dependence of the Seebeck coefficient. It turns out that the angular dependence is complex and strongly dependent on the TMR ratio. The angular dependence varies between a cosine-like dependence for a vanishing TMR ratio and a step-like function in the limit of an infinity TMR ratio. This variation explains the different experimental and theoretical observation of the angular dependence, because in theory the TMR ratio is at least one order of magnitude larger than in experiment. On the other hand it is also possible to deduct the size of the TMR ratio by analyzing the qualitative angular dependence of the TMS. Further, we showed that a TMS effect can exist even when no TMR effect is present.   

We acknowledge support from DFG SPP 1538 "SpinCaT". H.W.S. acknowledges support from EMR JRP SpinCal. 
We thank J. Langer and B. Ocker (SINGULUS Nano Deposition Technologies) for providing the MTJ stacks for the experiments.

%
%
%


\end{document}